# Bahamas: BAyesian inference with HAmiltonian Montecarlo for Astrophysical Stochastic background.

**Federico Pozzoli**  [1,2,¶], **Riccardo Buscicchio**  [2,3,4], **Antoine Klein**  [4], and **Daniele Chirico**  [2]

**1** Dipartimento di Scienza e Alta Tecnologia, Università dell'Insubria, via Valleggio 11, I-22100 Como, Italy  **2** Dipartimento di Fisica "G. Occhialini", Università degli Studi di Milano-Bicocca, Piazza della Scienza 3, 20126 Milano, Italy **3** INFN, Sezione di Milano-Bicocca, Piazza della Scienza 3, 20126 Milano, Italy **4** Institute for Gravitational Wave Astronomy & School of Physics and Astronomy, University of Birmingham, Birmingham, B15 2TT, UK **¶** Corresponding author

## Summary

The Laser Interferometer Space Antenna (LISA) (LISA Science Study Team, 2018) is a forthcoming space-based mission designed to detect gravitational waves (GWs). LISA consists of a constellation of three spacecraft arranged in a quasi-equilateral triangular configuration with an arm length of approximately $L \sim 2.5 \times 10^6 \mathrm{km}$. The constellation orbits the Sun while trailing the Earth. Each spacecraft is equipped with two telescopes and two lasers, enabling precise monitoring of the distances between test masses aboard each spacecraft. Given its design, LISA is sensitive to GW in the milli-hertz frequency band. In particular, LISA is expected to observe thousands of white dwarf (WD) binaries within the Milky Way, simultaneously, while the unresolved population of such binaries will overlap incoherently, forming the so-called Galactic foreground. One of the central challenges of the so-called global fit (Katz et al., 2025) is to jointly model both the resolvable and unresolvable WD populations. In particular, reconstructing the Galactic foreground is extremely difficult due to both computational and modeling complexities. In this article, we introduce bahamas, a tool designed to address some of these challenges from a global fit perspective. Additionally, we emphasize that accurately modeling the Galactic foreground also has applications in preliminary low-latency detection of massive black hole binaries (Cornish, 2022), which are compelling sources for multimessenger astronomy (Baker et al., 2019).

## Statement of need

The main idea behind the global fit algorithm is to use a Blocked Gibbs sampling technique to jointly analyze different GW sources, including stochastic backgrounds, instrumental noise, and the Galactic foreground. LISA is expected to sample data at $\sim 5s$, with a nominal mission duration of four years. This results in a large dataset for a full-band analysis of the stochastic components. Consequently, computational cost becomes a significant concern for the stochastic sector. Traditional sampling techniques, such as nested sampling or standard Markov Chain Monte Carlo (MCMC), might become prohibitively slow for this task. To address this issue, bahamas employs the No-U-Turn Sampler (NUTS) (Hoffman & Gelman, 2011), an adaptive variant of Hamiltonian Monte Carlo (HMC), which significantly enhances sampling efficiency. It uses the implementation provided by NumPyro (Phan et al., 2019), enablig automatic differentiation through JAX, while being agnostic on the hardware (i.e. CPU/GPU/TPU) architectures (Bradbury et al., 2018).

The reconstruction of the Galactic foreground is particularly challenging also due to its non-stationarity. Specifically, the Galactic foreground behaves as a cyclostationary process—a stochastic process with time-dependent periodic properties. This feature arises from the coupling between the highly anisotropic distribution of unresolved WDs in the Galaxy and the annually varying antenna pattern of LISA. The overlap of unresolved signals from a well-defined sky region results in a modulated stochastic signal in the time domain.

Similar to other works (Rosati & Littenberg, 2024), we use a Short-Time Fourier Transform likelihood rappresentation to analyze segmented data. While the chunking procedure mitigates the non-stationarity, it does not address variations in spectral amplitude within chunks caused by the modulation. (Digman & Cornish, 2022) proposes a phenomenological template that models the amplitude modulation as a superposition of sinusoidal forms. Instead, bahamas incorporates the modulation model proposed in (Buscicchio et al., 2024). The key advantage of this method consists in providing a modulation model that is both analytical and computationally efficient to evaluate, enabling simultaneous inference of both spectral parameters and sky distribution properties from the modulation.

## Software Description

The package includes two main command-line interfaces:

- `bahamas_data`: Data simulation and preprocessing.

- `bahamas_inference`: Parameter estimation and minimal diagnostics

Both scripts require two input files:

- `--config config.yaml`: Specifies the simulation and inference settings, sampler configuration, and output paths.

- `--sources sources.yaml`: Defines the sources to be injected and/or recovered. This includes the true physical parameters of the sources as well as the prior ranges used for inference.

The data consist of two datastreams—the A and E channels—which are specific combinations of Time-Delay Interferometry (TDI) variables (Tinto & Dhurandhar, 2021). In bahamas, data are generated in frequency domain, chunk by chunk. The duration of each chunk—and consequently the frequency resolution—can be configured via config.yaml. However, we recommend not using time lengths shorter than $10^4$s, which corresponds to a frequency resolution of approximately $\Delta f \sim 0.1$mHz, below which the characterization of LISA instrumental noise is not guaranteed. The noise model is defined by a two-parameter template that characterizes the amplitudes of two primary instrumental noise sources: the Test Mass (TM) noise and the Optical Metrology System (OMS) noise, both following predefined spectral shapes (European Space Agency (ESA), 2017).

The algorithm also allows for the analysis of stationary, isotropic, and Gaussian stochastic processes (e.g., a signal characterized by a power-law power spectral density), enabling the evaluation of the impact of multiple overlapping backgrounds and foregrounds.

We also provide the option to include data gaps, which represent periods during the mission when no useful data are available. These gaps can occur due to scheduled maintenance (scheduled gaps) or unforeseen hardware issues (unscheduled gaps). The goal of bahamas is not to mitigate the impact of these interruptions but rather to characterize their effect on the reconstruction of stochastic signals.

The algorithm is flexible to perform analyses with either full-resolution data or coarse-grained data over different chunks. In the former case, the likelihood describing the data follows a Whittle distribution (Moran & Whittle, 1951) in each segment, while in the latter, it degenerates to a Gamma distribution (Appourchaux, 2003) with degrees of freedom equal to the number of bins used in the averaging process.

## Performance

Below, we present a comparison of posterior probability reconstruction between HMC and nessai, both implemented in bahamas. In the example, we reconstruct the Galactic foreground spectrum and modulation alongside LISA instrumental noise. The computational cost between the two approaches may vary depending on inference settings. As a figure of merit, we consider a dataset corresponding to 6 months of mission duration, or 26 thousands (4 millions) effective datapoints for each Gamma (Whittle) likelihood evaluation. For the cyclostationary model inference over a 12-dimensional parameter space the hmc algorithm obtains 12 and 0.5 posterior samples per second for the Gamma and Whittle likelihood, while nessai does the equivalent with 2.6 (0.2) samples per second. While parallel chains in HMC are obtained independently, the number of simultaneous walkers in nested sampling affects significantly the performances. In this test we employed 10 cores and 16 cores for the 10 parallel HMC chains and the 16 parallel nested sampling walkers, respectively. Even if the speedup in using HMC is apparent from the metrics above, we highlight that Numpyro's performance, when internally parallelized over multiple chains, is known to be suboptimal and substantially dependent on the warmup chain length. In future release, we will provide code infrastructure to parallelize each chain production externally to the Numpyro API.

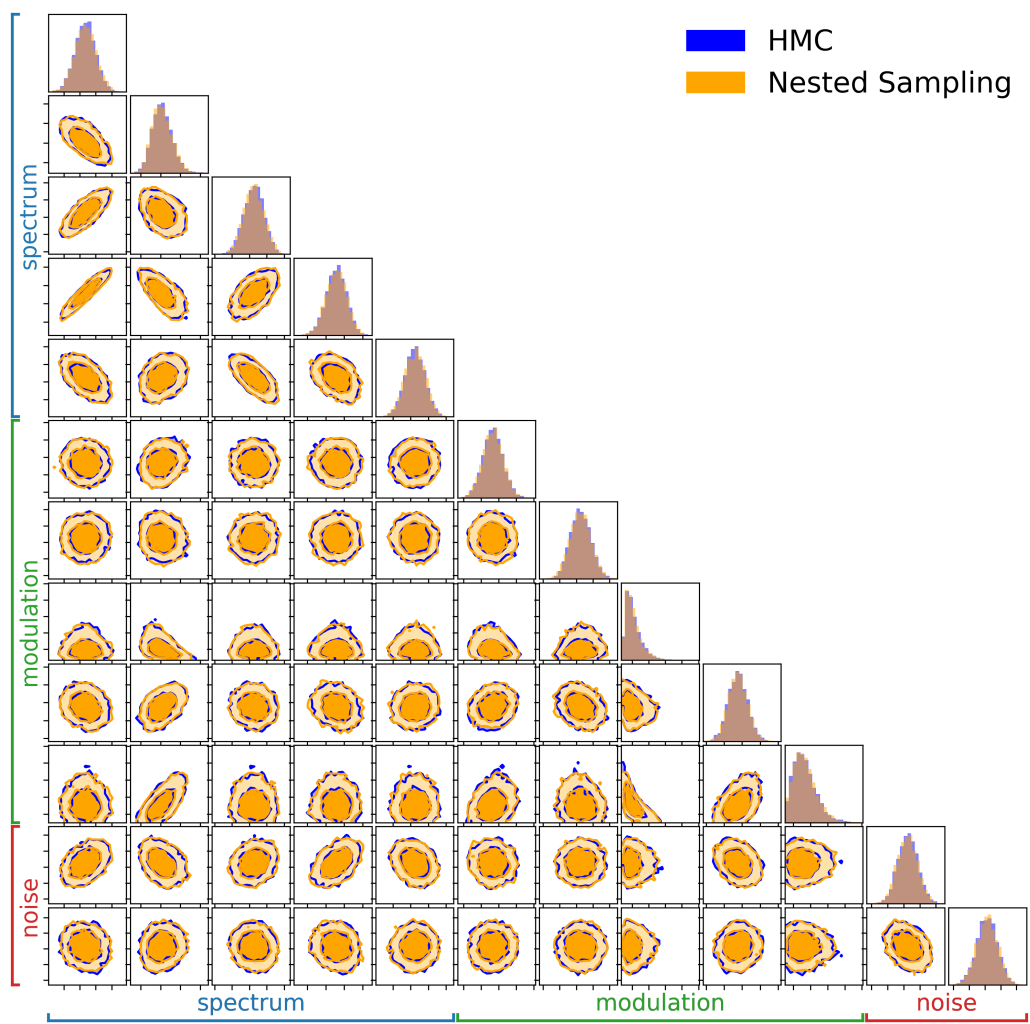

**Figure 1:** Posterior probability reconstruction of the spectrum and modulation for the Galactic foreground and LISA noise, as obtained using HMC and nested sampling.

## Outlooks

### Evidence

In future releases, we plan to include methods for computing the Bayesian evidence from HMC chains, enabling rigorous model selection. To estimate the evidence from HMC sample chains, viable techniques are thermodynamic integration and stepping-stone estimation (Maturana-Russel et al., 2019). Currently, bahamas also supports posterior probability exploration via nested sampling, using the nessai implementation (Williams et al., 2021), which provides evidence estimates as part of its output.

### Flexible Parametrization

Uncertainties in both the stochastic signal and the instrumental noise are expected for LISA, not only in their overall amplitude but also in their spectral shapes. For example, variations in the astrophysical modeling of white dwarf populations can lead to fluctuations in the shape of the Galactic foreground spectrum. Similarly, incorporating more realistic noise components can introduce additional complexity. To address these shape uncertainties, we plan to integrate the Expectation value of Gaussian Process (EGP) model, developed in (Pozzoli et al., 2024), as an example of a flexible parametrization.

### Other Non-stationarity

The cyclostationarity associated with the Galactic foreground is not the only source of non-stationarity in the LISA datastreams. Due to its actual orbit, LISA arm lengths will be unequal and vary over time. This effect introduces second-order non-stationarities in both the Galactic signal and the instrumental noise. At present, bahamas is not designed to address this issue, but further extensions of its capabilities are underway.

### TDI Correlations

Unequal arm length introduces also correlation between different TDI channels. These correlations can be accounted for in data analysis under the assumption of stationarity, as they appear as additional off-diagonal terms in the covariance matrix at each frequency (Hartwig et al., 2023). However, such correlations have not yet been explored or modeled for the Galactic foreground scenario. In future work, we plan to include a correlation matrix for stationary signal and noise and assess the impact of correlations in the non-stationary case.

## Acknowledgements

The authors thank A. Sesana, C. J. Moore, A.Vecchio, R.Meyer for useful comments. RB acknowledges support from the ICSC National Research Center funded by NextGenerationEU, and the Italian Space Agency grant Phase B2/C activity for LISA mission, Agreement n.2024-NAZ-0102/PER. Computational work was performed at Bicocca's Akatsuki cluster (B Massive funded).